# Antiferromagnetic VdW Phase at the Interface of Sputtered Topological Insulator/Ferromagnet-$Bi_2Te_3$/$Ni_{80}Fe_{20}$ Heterostructures

Nirjhar Bhattacharjee[1], Krishnamurthy Mahalingam[3], Adrian Fedorko[2], Valeria Lauter[4], Matthew Matzelle[2], Bahadur Singh[6], Alexander Grutter[5], Alexandria Will-Cole[1], Michael Page[3], Michael McConney[3], Robert Markiewicz[2], Arun Bansil[2], Donald Heiman[2], Nian Xiang Sun[1*]

[1]Northeastern University, Department of Electrical and Computer Engineering, Boston MA 02115, USA

[2]Northeastern University, Department of Physics, Boston MA 02115, USA

[3]Air Force Research Laboratory, Nano-electronic Materials Branch, Wright Patterson Air Force Base, OH 05433, USA

[4]Quantum Condensed Matter Division, Neutron Sciences Directorate, Oak Ridge National Laboratory, TN 37831, USA

[5]NIST Center for Neutron Research, National Institute of Standards and Technology, Gaithersburg, MD 20899, USA

[6]Tata Institute of Fundamental Research, Department of Condensed Matter Physics and Materials Science, Mumbai, 400005, India

**Abstract**

Magnetic ordering in topological insulators (TI) is crucial for breaking time-reversal symmetry (TRS) and thereby opening a gap in the topological surface states (TSSs) [1-6], which is the key for realizing useful topological properties such as the quantum anomalous Hall (QAH) effect, axion insulator state and the topological magnetoelectric effect. Combining TIs with magnetic materials can be expected to yield interfaces [26-28] with unique topological and magnetic phases but such interfaces largely remain unexplored. Here, we report the discovery of a novel antiferromagnetic (AFM) Van der Waals (VdW) phase at the interface of a sputtered *c-axis* oriented TI/FM ($Bi_2Te_3$/$Ni_{80}Fe_{20}$) heterostructure due to the formation of a Ni-intercalated $Bi_2Te_3$ VdW interfacial layer. The TI/FM heterostructure is shown to possess a significant spontaneous exchange bias and the presence of an AFM order at the interface via measurements of the hysteresis loop as well as the observation of compensated magnetic moments at the interface using polarized neutron reflectometry (PNR). An in-depth analysis of the structural and chemical properties of the interfacial AFM phase was carried out using selected area electron diffraction (SAED), electron energy loss spectroscopy (EELS), and X-ray photoelectron spectroscopy (XPS). These studies show evidence of solid-state reaction between the intercalated Ni atoms and $Bi_2Te_3$ layers and of the formation of topologically nontrivial magnetic VdW compounds. The Néel temperature of the interfacial AFM phase is 63 K, which is higher than that of typical magnetic topological insulators [53]. Our study shows how industrial CMOS-process-compatible sputtered TI/FM heterostructures can provide a novel materials platform for exploring the emergence of interfacial topological magnetic phases and high-temperature topological magnetic states.

**Introduction**

Introduction of magnetism in TIs to produce magnetic TIs (MTIs) has been achieved through magnetic doping of TIs [11-15], magnetic proximity effect [7-10], and synthesis of intrinsic magnetic compounds such as $MnBi_2Te_4$ that supports the QAH state [24] at ultra-low temperatures [16,17,20-25]. These exotic quantum states can help realize topological quantum computers and energy-efficient spintronic devices that can operate at temperatures close to room temperature. Van der Waals (VdW) materials are promising materials candidates in this connection due to the tunability and diversity of their topological properties.

MBE grown TI thin films have been shown to form interface phases when coupled with metals [26-28]. Interfaces of TI/FM heterostructures can lead to the emergence of new phases with exotic properties, their increased complexity notwithstanding [54]. Much of the existing experimental work has been performed on $Bi_2Se_3$ based systems. Te-based TIs are expected to possess an order of magnitude larger superexchange interaction strength compared to systems based on Se [16,17], making the latter more promising for creating topological magnetic phases. Notably, an AFM order in interfacial phases has not been previously reported.

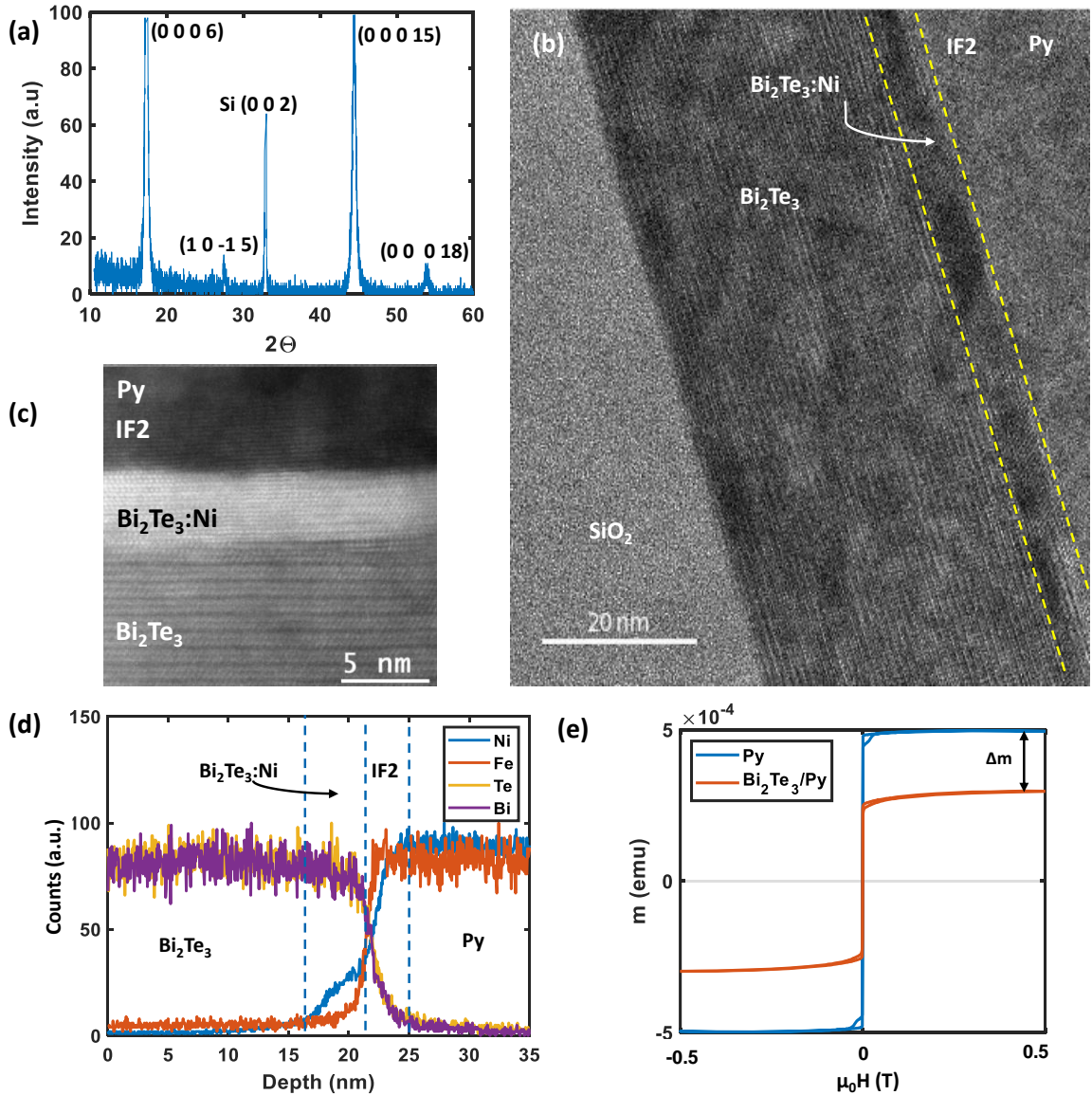

**Fig. 1** (**a**) XRD intensity vs 2θ showing *c*-axis orientation of $Bi_2Te_3$ film on Si substrate. (**b**) HRTEM image of sputtered $Bi_2Te_3$/Py heterostructure with interface layers labeled. (**c**) STEM-HAADF image of sputtered $Bi_2Te_3$/Py heterostructure showing clear atomically aligned VdW layered $Bi_2Te_3$:Ni interface formation. (**d**) Normalized concentration depth profile of elements in the heterostructure measured using EDS. (**e**) Room temperature m(H) loops showing loss of moments in the $Bi_2Te_3$/Py compared to control sample of Py with identical deposited Py thickness and film area. 1 emu = 1 mA m².

Sputter grown TIs with large spin-charge conversion efficiencies have been reported recently [30-34]. However, high-quality crystalline ordering in sputtered TIs has been elusive. This has prevented the study of many fascinating properties of interface phases and hindered the practical realization of spintronic devices based on robust topological properties that can be integrated in industrial CMOS production. Here, the discovery of a VdW antiferromagnetic (AFM) phase is reported in the interface of a highly $c$-axis oriented sputter grown TI, $Bi_2Te_3$, coupled with an FM metal alloy, $Ni_{80}Fe_{20}$ (Py). Diffusion of Ni from Py forms a VdW-layered, Ni-intercalated $Bi_2Te_3$ ($Bi_2Te_3$:Ni) phase with an in-plane easy axis AFM order. The presented sputter growth of crystalline textured TI ensures access to a topologically nontrivial surface which interacts with the Py layer, forming the $Bi_2Te_3$:Ni VdW AFM layer.

Measurements of the magnetic hysteresis loop show the appearance of a large spontaneous exchange bias (EB), signaling an AFM phase in proximity to the FM layer of Py with substantial exchange interaction between the AFM-FM layers [39-44]. The location of the AFM phase was identified using polarized neutron reflectometry (PNR) as a $Bi_2Te_3$:Ni layer formed at the interface of $Bi_2Te_3$/Py heterostructure. In addition, using selected area electron diffraction (SAED) we observed new diffraction peaks emerging in the AFM ordered interface, in addition to the $Bi_2Te_3$ peaks signalling the appearance of new planes along the crystalline $c$-axis. These new crystalline planes are formed due to the intercalation of Ni into $Bi_2Te_3$, and indicate the formation of Ni-based topological VdW compounds. Using electron energy loss spectroscopy (EELS) and X-ray photoelectron spectroscopy (XPS), we identified changes in electronic binding energies (BE) of Ni,Te and Bi [45-47] in the AFM $Bi_2Te_3$:Ni layer. These results demonstrate the formation of Ni-chalcogenides: Ni-Te bonds suggesting the formation of Ni-based VdW AFM topological compounds [16,17,49-52] in the $Bi_2Te_3$:Ni interface with a higher Néel transition temperature (~63 K) compared to MTIs reported in the literature [53].

**Results and Discussion**

**Crystal structure and interface morphology.** Heterostructures were grown by depositing 40 nm of c-axis oriented crystalline-textured $Bi_2Te_3$ on thermally oxidized $Si/SiO_2$ substrates using RF magnetron sputtering, followed by a 20 nm layer of Py (see Methods section for details). A reference sample was prepared with $Bi_2Te_3$ (40 nm)/$TiO_x$ (3 nm) for XRD analysis. The wide-angle XRD plot in Fig. 1(a) clearly shows a growth of $Bi_2Te_3$ with a significant crystalline orientation along the c-axis. The surface roughness of the $Bi_2Te_3$ layer was measured to be ~1 nm using X-ray reflectivity (XRR) (Supplementary Material Fig. S1) for the $Bi_2Te_3$/$TiO_x$ sample. This value of surface roughness is typical of sputter-grown thin films. These results show that high-quality, *c*-axis-oriented, crystalline-ordered TI can be grown using sputtering on an amorphous substrate, which enables us to study the interface phases in TI/FM heterostructures.

The morphology of the interface was studied extensively using several techniques. High-resolution transmission electron microscopy (HRTEM) was used to image the cross-section of the $Bi_2Te_3$/Py heterostructure as shown in Fig. 1 (b). The HRTEM and scanning transmission electron microscopy-high angle annular diffraction (STEM-HAADF) images in Figs. 1(b,c) clearly show VdW layers, which confirms the crystalline-textured, *c*-axis-oriented growth of the $Bi_2Te_3$ thin film. Energy dispersive X-ray spectroscopy (EDS) was used to characterize the elemental composition of the heterostructure as presented in Table 1 (details of analysis in Supplementary Materials Fig. S4). The obtained Bi:Te ratio of 2:3 confirms a stoichiometric composition of the sputtered $Bi_2Te_3$ thin film. It is important to note that a significant diffusion of Ni occurs within the interface, as illustrated in the normalized EDS depth profile in Fig. 1(d). This diffusion of Ni is accompanied by a substantial loss of magnetic moment, **m**, by ~40% compared to a control sample of Py as shown in the room temperature m(H) loop in Fig. 1(d). The diffused Ni intercalates in the $Bi_2Te_3$ VdW gaps and undergoes solid-state reaction, possibly catalyzed by the topological surface state (TSS) electrons [27,29]. A second interface layer forms because of depletion of Ni, wherein trace amounts of Bi and Te are also detected, which is referred to as the IF2 layer. The STEM-HAADF image suggests that the 5 nm thick $Bi_2Te_3$:Ni AFM layer is a VdW layered phase which will be further verified using various techniques.

**Table 1. Elemental composition in percentage of the $Bi_2Te_3$/$Bi_2Te_3$:Ni/IF2/Py heterostructure shown in Fig. 1.**

| Element | $Bi_2Te_3$ | $Bi_2Te_3$:Ni | IF2 | Py |
|---|---|---|---|---|
| Bi | 39 | 22 | 3 | 0.03 |
| Te | 60.28 | 34.31 | 5.24 | 0.20 |
| Ni | 0.38 | 39.38 | 69.90 | 80.79 |
| Fe | 0.31 | 4.11 | 21.65 | 18.99 |

**Magnetic properties.** To study the magnetic properties in the sputtered $Bi_2Te_3$/$Bi_2Te_3$:Ni/IF2/Py

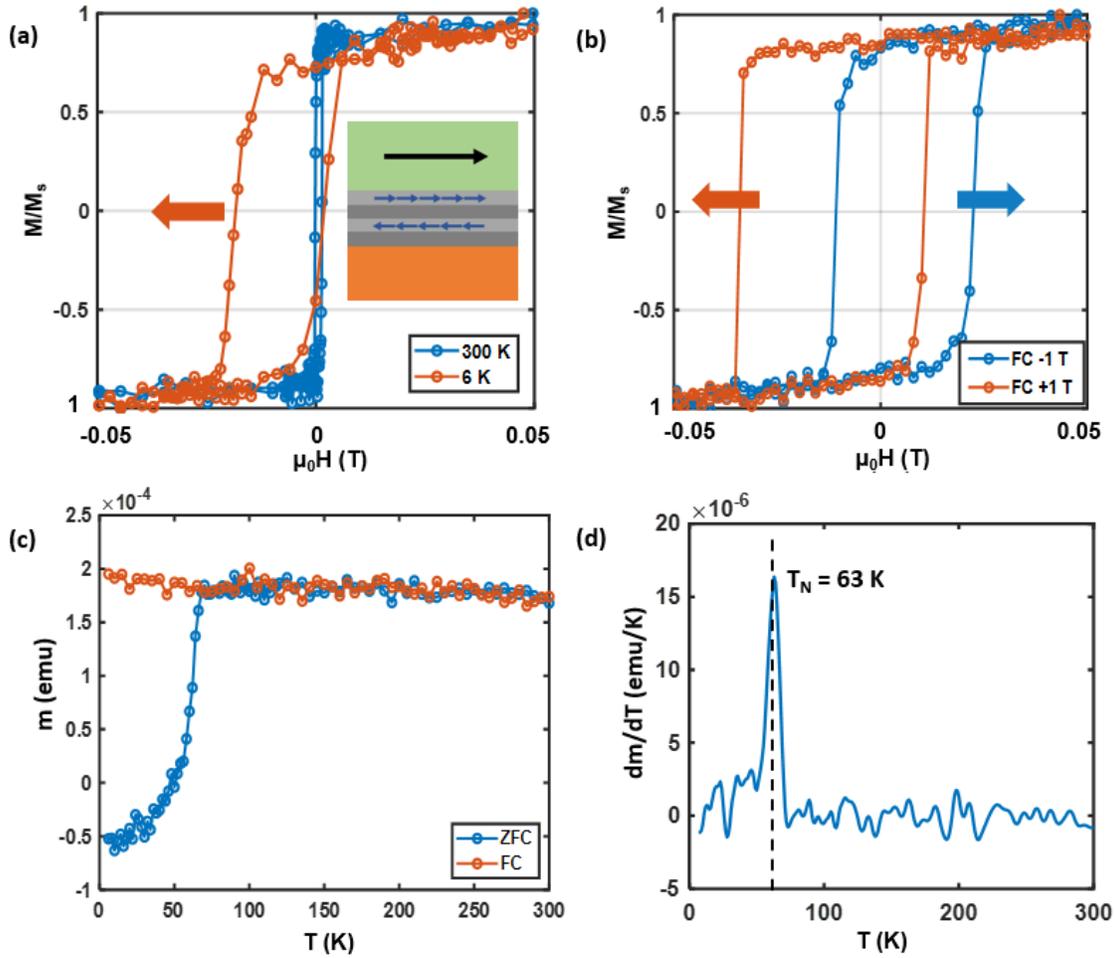

**Fig 2 (a)** ZFC hysteresis loops at 300 K and 6 K showing exchange bias at low temperature. **(b)** Hysteresis loops for FC at -1 T and +1 T showing switching of EB. **(c)** ZFC and FC m(T) plots measured at 5 mT bias field. **(d)** Derivative, dm/dT, of the ZFC plot in **(c)** showing a peak at ~63 K corresponding to the Néel temperature of the heterostructure, m(H) loop measurements were carried out with the field in the plane of the sample with a superconducting quantum interference device (SQUID) magnetometer. As expected, the in-plane hysteresis loops measured at temperature, T = 300 K are centered with a small coercive field, $\mu_0 H_c \approx 0.5$ mT, which is comparable to the Py control sample in Fig. 1(e). In contrast, the zero-field-cooled (ZFC) and field-cooled (FC) low-temperature measurements exhibit a sizable spontaneous EB along with a significant enhancement of $\mu_0 H_c$ as shown in Fig 2(a). This EB shift is a characteristic of presence of a large exchange interaction strength in FM/AFM interfaces [40-42]. The ZFC hysteresis loop measured at 6 K is shifted by EB ≈ 8 mT to the left and an enhanced coercive field which was determined to be $\mu_o H_c$

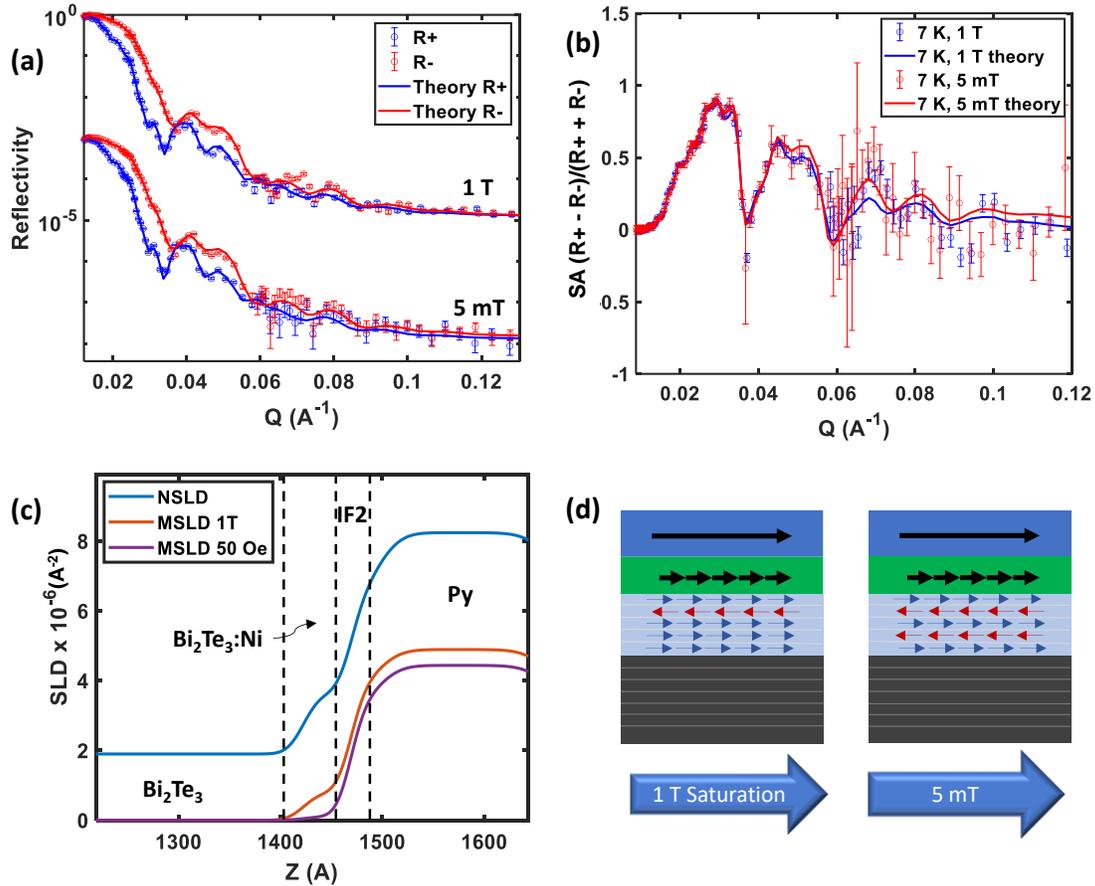

**Fig. 3 (a)** PNR reflectivity plots for measurements at 1 T and 5 mT with fits for the theoretical model. (**b**) SA plots derived from (a) for raw data and model fitting. (**c**) MSLD and NSLD depth profiles of showing AFM phase emerging in the Ni-rich IF1 interfacial layer. (**d**) Schematic illustrations of orientation of magnetic moments across the layers (bottom to top: VdW $Bi_2Te_3$ in black, VdW $Bi_2Te_3$:Ni in light blue, IF2 in green, Py in deep blue) at 1 T saturation field and 5 mT low field. All these measurements were performed at a temperature of 7 K (Not to scale).

~9.5 mT. FC hysteresis loops were also measured where the samples were cooled with an applied field of ±1 T. As shown in Fig. 2(b), EB field switched from ~−18.0 mT to ~+18.0 mT for FC of +1 T and −1 T, respectively. This confirms the presence of a large exchange coupling that exists between the Py FM and the Bi$_2$Te$_3$:Ni AFM layers, with a large in-plane magnetic anisotropy [44] indicative of an A-type AFM [16,17] material (more evidence of A-type AFM in Supplementary Materials Sec. 8). The Néel temperature at which the AFM phase emerges was obtained from m(T) measurements performed in both FC and ZFC conditions using a constant applied field of 5 mT as shown in Figs. 2 (c,d). Figure 2(d) plots the derivative, dm/dT of the ZFC m(T), showing a sharp peak at ~63 K, which is the Néel transition temperature of the AFM phase [41,42]. This Néel transition temperature of the Bi$_2$Te$_3$:Ni AFM phase is higher than transition temperatures of MTIs reported in the literature [53].

The location of the layer with AFM ordering present in the heterostructure was investigated by depth-sensitive PNR measurements [55, 56] at the Oak Ridge National Laboratory (ORNL) Spallation Neutron Source using the MagRef reflectometer [57]. Experiments were performed at T = 7 K under ZFC conditions using 1 T and 5 mT applied fields. The theoretical model was fit to the PNR profiles using Refl1D software for the reflectivity and spin-asymmetry experimental data, $SA = (R^+ - R^-)/(R^+ + R^-)$ as shown in Fig. 3 (a,b). The structural and magnetic depth profiles were parameterized using the nuclear and magnetic scattering length densities (NSLD and MSLD, respectively) as shown in Fig 3(c) (full PNR profile in Supplementary Materials Fig. S7). The NSLD and MSLD depth profiles obtained from the fit to the reflectivity represent the structural and magnetic depth profiles of the heterostructure respectively. Note that the reduction in NSLD by $\sim 2 \times 10^{-6}$ Å$^{-2}$ in the IF2 layer is compensated by a gain in NSLD by approximately the same value in the Bi$_2$Te$_3$:Ni layer. This suggests that roughly the Ni lost by the Py in the IF2 layer forms the distinct Bi$_2$Te$_3$:Ni interfacial layer, which is in agreement with the STEM-HAADF image in Fig. 1(c). For the applied bias field of 1 T, most of the moment vectors are aligned along the field direction, including the ones in the interfacial layer, illustrated in Fig. 3(d). But at a low field of 5 mT, the AFM order in the Bi$_2$Te$_3$:Ni VdW layers yields largely compensated moments. From the MSLD depth profile in Fig. 3(c), the Bi2Te3:Ni layer was observed to have a much larger value of 0.26×10$^{-6}$ Å$^{-2}$ at 1 T compared to 0.034×10$^{-6}$ Å$^{-2}$ at 5 mT. The lower value of moments by an order of magnitude at 5 mT verifies the AFM ordering because of Ni-intercalating in VdW gaps and reacts with Bi$_2$Te$_3$. A lower value of MSLD can also mean that the moments are oriented out of the film plane, as

PNR is insensitive to perpendicular component of moments. However, the large exchange bias for in-plane m(H) loop measurements as shown in Fig. 2 and significantly low remnant magnetization in out-of-plane m(H) loop measurement (Supplementary Materials, Fig. S7) provides clear evidence of largely in-plane easy-axis AFM order in the $Bi_2Te_3$:Ni layer with small canting of moments out-of-plane.

**Structural and Chemical Properties.** The *c*-axis crystalline-oriented texture of the VdW-layered $Bi_2Te_3$ and the $Bi_2Te_3$:Ni AFM layers are clearly identified in the HRTEM and STEM-HAADF images in Fig. 1(b,c) and SAED patterns in Figs. 4 (a-c). In order to understand the structural and chemical properties of the AFM $Bi_2Te_3$:Ni layer, SAED followed by cross-section EELS and depth-dependent XPS measurements were performed. These measurements provide strong evidence of formation of topologically nontrivial VdW compounds in the AFM interfacial $Bi_2Te_3$:Ni layer as a result of solid-state reaction between intercalated Ni and $Bi_2Te_3$.

SAED measurements firmly demonstrate a VdW-layered structure in the AFM $Bi_2Te_3$:Ni layer as shown by the single line of diffraction spots in Fig. 4(a), as well as in the $Bi_2Te_3$ of Fig. 4(b). Figure 4(c) plots the diffraction intensities, where additional peaks emerge in the AFM $Bi_2Te_3$:Ni layer compared to the peaks representing quintuple layers (QL) of $Bi_2Te_3$. These additional diffraction peaks confirm intercalation of Ni in $Bi_2Te_3$ and formation of new crystalline planes parallel to the crystalline *c*-axis. A similar pattern of the SAED peaks in $Bi_2Te_3$:Ni layer compared to the $Bi_2Te_3$ layer further confirms the VdW nature of the layer. Of particular interest is the composition of the $Bi_2Te_3$:Ni layer. The presence of topological compounds, such as NiTe, $NiTe_2$, $NiBi_2Te_4$ and $(Ni,Bi)_2Te_3$ are expected by qualitative comparison of the diffraction peaks with the possible reaction products of Ni and $Bi_2Te_3$ (Supplementary Materials Fig S9). These key pieces of information are strong indications of the formation of a Ni-based VdW-layered magnetic topological material phase in the AFM $Bi_2Te_3$:Ni layer.

Further evidence supporting the formation of topologically nontrivial magnetic compounds in the AFM VdW $Bi_2Te_3$:Ni layer is provided by cross-sectional EELS and depth-dependent XPS measurements, shown in Fig 4(d). The EELS measurements for core shell electrons were performed on the layers marked on the STEM-HAADF cross-section images shown in Fig. 1(c). For the $Bi_2Te_3$:Ni AFM layer, Fig. 4(d) shows new pre-edge features emerging at ~30 eV lower binding energies (BE) prior to Ni and Fe L-shell edges, which were absent for the Py reference layer. In addition to these pre-edge features, a change in L3:L2 peak ratio of Ni from 1.90 in the Py layer to 1.41 in the $Bi_2Te_3$:Ni AFM layer was observed. It is concluded that the emergence of the pre-edge features and reduction in L3:L2 peaks ratio can be attributed to the increase in valence state of Ni from a metallic state in Py towards $Ni^{2+}$ oxidation state in the $Bi_2Te_3$:Ni layer [45,46]. It is also noted that the Fe L-shell peaks showed reduction in L3:L2 ratio from 1.94 in the Py region to 1.25 in the $Bi_2Te_3$:Ni layer. However, Fe in the $Bi_2Te_3$:Ni layer has a much lower concentration of ~4.1% compared to ~39.4% for Ni as determined by EDX. Hence, the AFM phase should be predominantly attributed to Ni-based compounds in the $Bi_2Te_3$:Ni layer.

The presence of topological magnetic compounds in the AFM $Bi_2Te_3$:Ni layer is also signaled by depth-dependent XPS measurements, indicating Ni-Te bonding. Figure 4(e) compares the normalized XPS spectra of Ni in the Py reference and the $Bi_2Te_3$:Ni AFM layers. The satellite peak at 858.7 eV shifts by ~1 eV to higher BE, which is a signature of Ni-chalcogenide (Ni-Te) bonds [47]. Similarly, a large shift of ~1 eV is also observed in the Fe 2p3 binding energies (Supplementary Materials Fig. S6). Further, as shown in Figs. 4 (g,h) and Table 2, the Te 3d5 and Bi 4f peaks also experience a shift of ~0.5 eV from 572.8 eV and ~0.3 eV from 157.4 eV (main peak positions) towards higher and lower binding energies, respectively, in the $Bi_2Te_3$:Ni layer compared to the reference layer of $Bi_2Te_3$. The shift in XPS peaks for Ni, Te and Bi suggest the formation of $Ni_xBi_yTe_z$ compounds. Finally, all these results, from XPS along with SAED and EELS characterisations, provide strong evidence for formation of new VdW compounds with Ni-Te bonds due to the intercalation of Ni in $Bi_2Te_3$ in VdW gaps.

Table 2. XPS peak positions of interest for Ni, Fe, Bi and Te for $Bi_2Te_3$:Ni layer compared to reference Py and $Bi_2Te_3$ layers

| Element | Py | Bi2Te3:Ni | Bi2Te3 | ΔE |
|---|---|---|---|---|
| Ni | 858.7 | 859.6 | - | 0.9 |
| Fe | 712.1 | 713.0 | - | 0.9 |
| Te | - | 572.8 | 572.3 | 0.5 |
| Bi | - | 157.1 | 157.4 | -0.3 |

Theoretical calculations of $NiBi_2Te_4$, which belongs to the $MBi_2Te_4$ (M = Mn, Ni, V, Eu etc.) family, has been reported to be an intrinsic MTI compound with a large exchange energy and in-plane magnetic anisotropy [16,17]. With this in mind, the emergence of a large exchange-bias from the AFM-ordered $Bi_2Te_3$:Ni VdW layer points towards the emergence of Ni-based AFM phases with topologically nontrivial properties. Qualitative comparison of the positions of the *c*-axis-oriented SAED peaks with theoretical diffraction peak positions of possible $Ni_xTe_yBi_z$ VdW materials, raises the possibility of the presence of topologically nontrivial compounds such as $NiBi_2Te_4$, $(Ni,Bi)_2Te_3$, NiTe and $NiTe_2$ in the $Bi_2Te_3$:Ni AFM layer [16,17,51]. Of these however, NiTe and $NiTe_2$ are known to be paramagnetic [47]. This makes the intrinsic MTI compounds, $NiBi_2Te_4$ (intercalation of Ni) and $(Ni,Bi)_2Te_3$ (substitution of of Bi sites by Ni) a highly likely candidate for the AFM ordering in the $Bi_2Te_3$:Ni layer. In such VdW-layered AFM systems, the magnetic ordering is usually dictated by superexchange interactions [16,17]. The interstitial magnetic Ni atoms are believed to be FM ordered within the VdW plane, but each plane is coupled by RKKY exchange in the out of plane direction, giving rise to an A-type AFM ordering [16,17, 48-50]. Of further interest is high Néel transition temperature of ~63 K measured for the $Bi_2Te_3$:Ni AFM layer, which is higher than the MTIs [53] reported in literature. This makes the present $Bi_2Te_3$:Ni phase an appealing candidate for high-temperature QAH and other topological phases that are required for realizing energy efficient spintronics and topological quantum computing applications.

**Conclusion**

Sputtered TI/FM heterostructures of highly *c*-axis-oriented $Bi_2Te_3$ coupled with the FM alloy Py were shown to exhibit a large exchange bias field in Ni-intercalated Bi2Te3, forming a clear AFM VdW layer. The $Bi_2Te_3/Ni_{80}Fe_{20}$ heterostructures were grown on amorphous thermally oxidized Si substrates. Using PNR, the AFM ordering was confirmed in the Ni-intercalated $Bi_2Te_3$ VdW layer. The intercalated Ni undergoes a solid-state reaction with $Bi_2Te_3$, which is believed to be catalysed by delocalized TSS

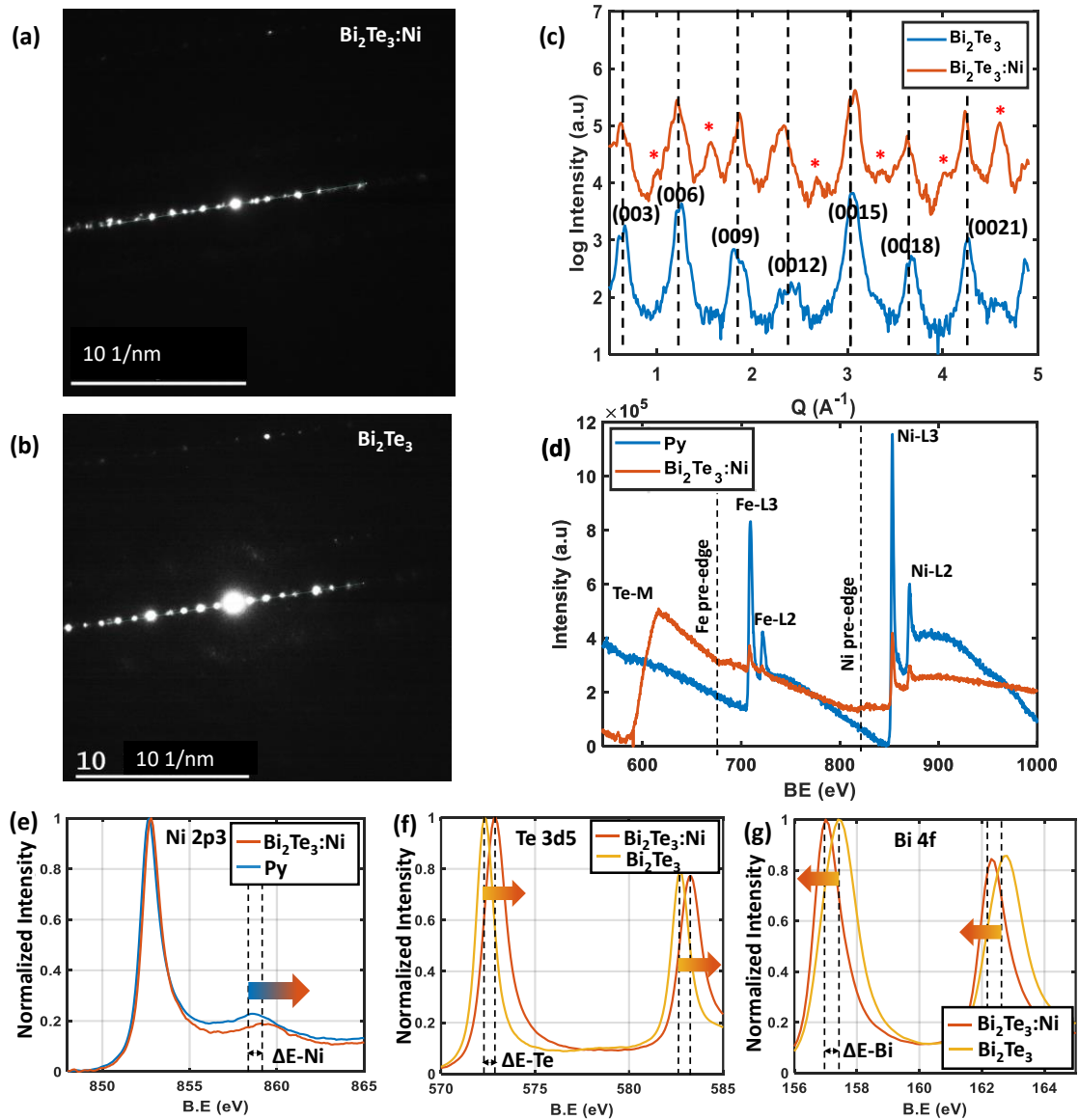

**Fig. 4.** SAED intensity for sputtered $Bi_2Te_3$/Py from STEM-HAADF image in Fig. 1c showing diffraction peaks at (**a**) $Bi_2Te_3$:Ni and (**b**) $Bi_2Te_3$ regions, (**c**) SAED Peak positions vs. $d^{-1}$ extracted from (a,b) to identify new peaks emerging in the $Bi_2Te_3$:Ni layer. (**d**) EELS spectra for Te M-shell and Fe, Ni L-shell electrons corresponding to sputtered $Bi_2Te_3$/Py heterostructure interface layer cross section with new pre-edges and reduced L3:L2 ratios for Ni and Fe in the $Bi_2Te_3$:Ni region compared to Py reference region corresponding to STEM-HAADF image in Fig. 1(c). Normalized XPS spectra for: (**e**) Ni 2p3/2 showing shift in satellite peak by ~1 eV towards higher BE; (**f**) Te 3d5/2 showing ~0.5 eV shift towards higher BE and (**g**) Bi 4f showing ~0.3 eV shift towards lower BE.

Bi$_2$Te$_3$ peaks, confirming intercalation of Ni in Bi$_2$Te$_3$ VdW gaps and the formation of additional planes oriented along the crystalline *c*-axis. Furthermore, using cross-section EELS and depth-dependent XPS measurements, spectral changes were observed for Ni, Fe, Bi, and Te in the Ni-intercalated Bi$_2$Te$_3$ AFM layer, as compared to the reference Bi$_2$Te$_3$ and Py layers. These are attributed to the formation of Ni-Te bonds in the interface layer. Qualitative comparison of SAED peak positions and the detection of Ni-Te bonds in the Ni-intercalated Bi$_2$Te$_3$ AFM phase are strong indications of *c*-axis-oriented Ni$_x$Te$_y$Bi$_z$ VdW compounds that are known to be topologically nontrivial. Furthermore, the 63 K Néel transition temperature is considerably higher than the magnetic transition temperatures of recent experimentally synthsized MTIs, which makes the present interfacial phases in sputtered TI/FM heterostructures potential candidates for high-temperature QAH and other promising topological materials. These results open the path for further exploration of industrial CMOS compatible sputter-grown TIs and TI/FM material systems for high-temperature topological material systems and realization of energy-efficient topological spintronic devices.

**Methods**

**Material Growth.** Heterostructures of 30 nm Bi$_2$Te$_3$ were grown by co-sputtering a composite Bi$_2$Te$_3$ target with a Te target using RF magnetron sputtering at 90 W and 20 W, respectively, with 4 mtorr (0.53 Pa) Ar pressure on thermally oxidized Si substrates. The base pressure of the sputtering chamber was <2×10$^{-7}$ Torr (2.67×10$^{-5}$ Pa). The samples were grown with substrates maintained at 250°C. Samples were further annealed inside the chamber at 45 mTorr (6 Pa) Ar pressure for 25 minutes at 250°C to achieve high-quality crystalline c-axis oriented textured growth. The samples were cooled to room temperature in a high vacuum for ~5 hours before deposition of other layers. For the magnetic sampless, 20 nm Py and 3 nm Ti capping layer were deposited at room temperature after deposition of Bi$_2$Te$_3$. The Ti layer subsequently oxidized to TiO$_x$ on exposure to atmosphere.

**XRD Characterisation.** A nearly perfect collimated and background-free beam of CuKα1 radiation (wavelength λ = 1.54056 Å) is impinged on the sample surface and X-ray scattering intensity was collected by a two-dimensional charged-coupled device (CCD). The sample alignments are done on Bi$_2$Te$_3$ layer reflections and scans are performed along the Bi$_2$Te$_3$ growth direction. The Bragg reflections are indexed according to the Bi$_2$Te$_3$ bulk hexagonal unit cell. The x-axis in Fig. 1(a) is indexed in terms of the hexagonal unit cell of the Bi$_2$Te$_3$, as indicated by ($h, k, -(h+k), l$) where $h$, $k$, and $l$ are the Miller indices. A highly c-axis oriented growth (along ($0,0,3l$) direction) of Bi$_2$Te$_3$ was identified using XRD.

**TEM, EELS and XEDS Characterisation.** Samples for TEM investigations were prepared by focused ion beam milling (FIB) using a Ga+ ion source. Prior to TEM observation an additional cleaning procedure was performed by Ar-ion milling to reduce surface amorphous layer and residual Ga due to the FIB process. The TEM observations were performed using a Talos 200-FX (ThermoFiszher Scientific Inc) TEM operated at an acceleration voltage of 200kV. XEDS measurements were performed using the ChemiSTEM (ThermoFisher Scientific) technology and acquisition and processing of the spectra was performed by spectrum imaging technique using the Esprit 1.9 (Brucker Inc.) software. The Nanobeam diffraction and STEM/EELS studies were performed using an aberration-corrected (image) Titan operated at an accelerating voltage of 200kV. The EELS data acquisition was performed using a GIF-Quantum (Gatan, Inc) spectrometer and processed using the DigitalMicrograph 2.10 (Gatan, Inc.) software.

**Hysteresis Loop Measurements.** Magnetization m(H) and m(T) measurements were obtained using a Quantum Design superconducting quantum interference device (SQUID) magnetometer. Hystersis loop m(H) measurements were carried out at various temperatures between 6 K and 300 K. The ZFC and FC m(T) measurements were obtained while increasing the temperature in an applied field of 50 Oe (1 Oe = 79.6 A/m), and FC measurements were performed after cooling the sample under an applied field of 1 T. Room temperature m(H) measurements in Fig. 1 (e) were taken using a vibrating sample magnetometer (VSM).

**PNR Characterisation.** PNR experiments were carried out on the Magnetism Reflectometer at the Spallation Neutron Source at Oak Ridge National Laboratory [57]. A neutron beam with a wavelength band of 2.6–8.6 Å and a high polarization of 98.5 % was used. Measurements were performed in a closed cycle refrigerator in an applied external magnetic field using a Bruker electromagnet with a maximum magnetic field of 1 T. In the time-of-flight method, a collimated polychromatic beam of polarized neutrons with a wavelength band $\Delta\lambda$ impinges on the film at a grazing incidence angle $\theta$. In the film it interacts with atomic nuclei and the spins of unpaired electrons [58, 59]. The reflected intensity is measured as a function of wave vector transfer, $Q = 4\pi\sin(\theta)/\lambda$, for two neutron polarizations $R+$ and $R-$, with the neutron spin parallel (+) or antiparallel (−) to the direction of the external field, $H_{ext}$. To separate nuclear from magnetic scattering, the data are presented in the form of the spin–asymmetry ratio SA = $(R+ - R-)/(R+ + R-)$ as shown in Figs. 3(b) and S7(b), where SA = 0 means there no magnetic moment in the system. Electrically neutral, spin-polarized neutrons penetrate the entire structure of the film and probe depth profile of magnetic and structural composition of the film interfaces down to the substrate with a resolution of 0.5 nm. The depth profiles of the nuclear and magnetic scattering length densities (NSLD and MSLD) correspond to the depth profile of the chemical and in-plane magnetization vector distributions on the atomic scale, respectively [57-59]. Based on these neutron scattering merits, PNR serves as the powerful technique to simultaneously and nondestructively characterize chemical and magnetic nature of buried interfaces [57]. Neutron scattering measurements were performed on a 2×2 cm$^2$ surface samples.

**XPS Depth Profile.** Chemical composition of the surface was characterized using a PHI Versaprobe II X-ray photoelectron spectrometer with a scanning monochromated Al source (1486.6 eV, 100 W, spot size 200 μm). Depth profiling was accomplished using the instrument's C60+ ion source. The takeoff angle between the sample surface and analyzer was 45º, and the X-ray beam collected Ni2p, Fe2p, Te3d, Bi4f, and Si2p elemental information while rastering over a 200 x 1400 µm2 area. Sputtering occurred in 1 min intervals, while the sample was moved using concentric Zalar rotation at 1 rpm. The C60+ source was operated at 1 kV and 0.5 µA and rastered over a 2x 2 mm2 area at an angle 70º to the surface normal. Valence states of elements were determined by comparing the shift in XPS peaks in BE, and the relative sensitivity factors were provided in PHI's Multipak processing software. All data were background-subtracted and smoothed using a five-point quadratic Savitzky-Golay algorithm. The relative position of the layers were based on amplitude of the spectra and relative shift in peak positions. Spectra peaks were fit in CasaXPS, and data were plotted and analyzed using Matlab.

**Acknowledgement**

This work is partially supported by the U.S Army under grant no. W911NF20P0009, the NIH Award UF1NS107694 and by the NSF TANMS ERC Award 1160504. The work of A.B. and M.M was supported by the US Department of Energy (DOE), Office of Science, Basic Energy Sciences Grant No. DE-SC0019275 and of R.M. by DOE grant number DE-FG02-07ER46352, and benefited from Northeastern University's Advanced Scientific Computation Center and the Discovery Cluster, and the National Energy Research Scientific Computing Center through DOE Grant No. DE-AC02-05CH11231. A portion of this research used resources at the Spallation Neutron Source, a DOE Office of Science User Facility operated by the Oak Ridge National Laboratory. We thank Charles Settens and Libby Shaw (MIT, Materials Research Laboratory) for help with XRD and XPS measurements. Certain commercial equipments are identified in this paper to foster understanding.  Such identification does not imply recommendation or endorsement by Northeastern University, AFRL, ORNL, NIST and TIFR.

**Table 1. Elemental composition in percentage of the $Bi_2Te_3/Bi_2Te_3:Ni/IF2/Py$ heterostructure shown in Fig. 1.**

| Element | $Bi_2Te_3$ | $Bi_2Te_3$:Ni | IF2 | Py |
|---|---|---|---|---|
| Bi | 39 | 22 | 3 | 0.03 |
| Te | 60.28 | 34.31 | 5.24 | 0.20 |
| Ni | 0.38 | 39.38 | 69.90 | 80.79 |
| Fe | 0.31 | 4.11 | 21.65 | 18.99 |